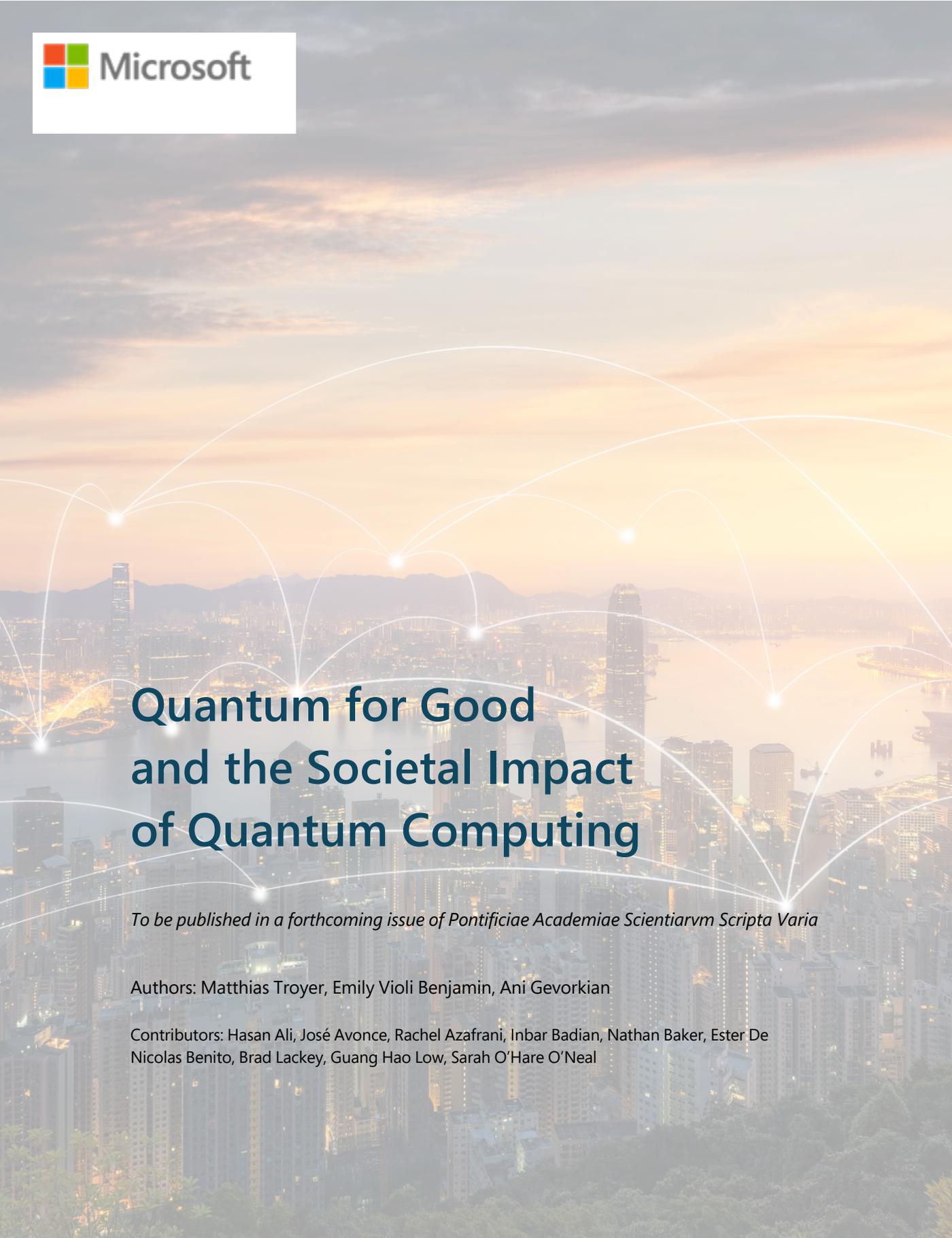

# Quantum for Good and the Societal Impact of Quantum Computing

Authors: Matthias Troyer, Emily Violi Benjamin, Ani Gevorkian

Contributors: Hasan Ali, José Avonce, Rachel Azafrani, Inbar Badian, Nathan Baker, Ester De Nicolas Benito, Brad Lackey, Guang Hao Low, Sarah O'Hare O'Neal

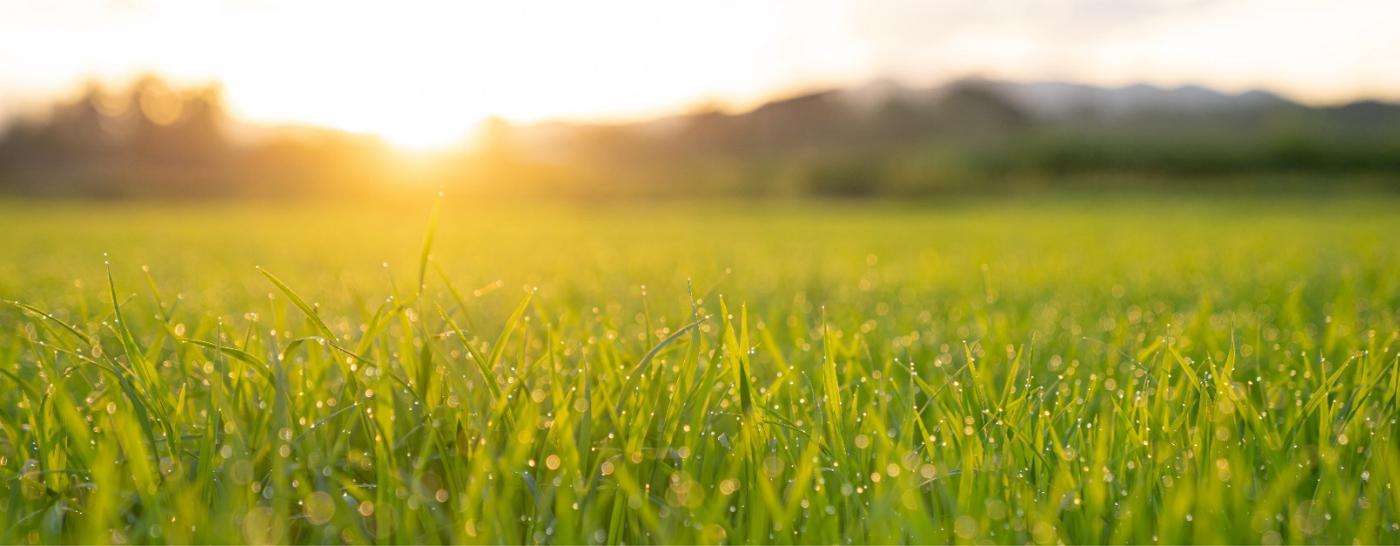

Quantum computing promises to help humanity solve problems that would otherwise be intractable on classical computers. Unlike today's machines, quantum computers use a novel computing process that leverages the foundational quantum mechanical laws of nature. This unlocks unparalleled compute power for certain applications and promises to help solve some of our generation's gravest challenges, including the climate crisis, food insecurity, and widespread disease.

No one entity will be able to realize this end state alone. Developing a fault-tolerant quantum supercomputer and a vibrant ecosystem around it will require deep partnerships between industry, governments, and academia. It will also require collective action to enable and promote positive applications of quantum computing and ensure that the safe and responsible use of the technology is at the center of its development and deployment. Achieving these objectives will require focusing on three priorities:

1. **Impact.** Ensure quantum computing benefits all of humankind by developing quantum solutions to solve critical, global problems.

2. **Use.** Protect against malicious use by accelerating the deployment of quantum-safe cryptography and developing governance processes and controls for the responsible use of quantum machines.

3. **Access.** Democratize the potential for economic growth across all of society through skilling, workforce and ecosystem development, and digital infrastructure.

This paper discusses each in turn.

# Impact: Ensure quantum computing benefits all of humankind

## Practical quantum advantage: the impactful problems for a quantum computer

Quantum computers offer the potential to solve complex problems that are intractable using classical computers. This has led to increasingly unchecked expectations and hype. A plethora of difficult problems have been postulated to benefit from quantum computing: from traditionally understood cryptanalysis and chemistry and materials science applications to database search, weather and stock market predictions, optimization, machine learning, protein folding, drug design and more. There are quantum algorithms with asymptotic quantum speedup for many of these problems, but where will this lead to a practical quantum advantage with reasonable execution times?

The reality is that quantum computing will neither accelerate the solution of every problem nor replace classical computers. Operations on a quantum computer, especially fault-tolerant ones, are many orders of magnitude more costly than classical digital logic gates that require only a few transistor switches. For practical quantum advantage, which we define as quantum computers with the ability to outperform classical ones in hours or days instead of years or longer [1], we must focus on small-data problems with super-polynomial quantum speedup [2], paying special consideration to input/output bottlenecks.

Considering these realities, the most effective applications for quantum computers [1], besides cryptanalysis using Shor's algorithm [3], are simulations of quantum systems [4], especially applied to chemistry, biochemistry, and materials science. Even if no other practical quantum applications are discovered, the impact of quantum in these fields cannot be understated: ninety-six percent of all manufactured goods rely on chemistry or materials science [5].

At scale, quantum computing will enable significantly enhanced predictive accuracy in quantum simulations by overcoming existing accuracy challenges in modeling electronic structures. These accuracy challenges arise from approximations required to model systems with large numbers of correlated electrons. Complete active space configuration interaction methods are desired to accurately model correlated electron behavior; however, the classical computational cost for these methods scales exponentially with problem size. In the future, we expect that scaled fault-tolerant quantum computers will offer an exponential speed-up for modeling correlated electron behavior with higher accuracy than is possible classically.

This capability will allow for more reliable computational understanding and design of materials and chemicals. The first applications may be modeling catalytic systems [6], such as efficient catalysts for carbon fixation to help combat global warming, more sustainable fertilizer production, and cleaner combustion. Longer term applications will include correlated quantum materials, especially those containing transition metal and rare earth elements, enabling applications such as more efficient batteries and solar cells, higher temperature superconductors, and more [7,8]. Advances in these fields map directly to chemistry and material science challenges identified in the United Nations (U.N.) Sustainable Development Goals for 2030 [9].

Of those applications, let us focus on computational catalysis—specifically, the simulation of catalytic reaction processes [6,10]. Catalysts accelerate chemical reactions and are the foundation of the chemical industry. Understanding reaction mechanisms and outcomes requires calculating the reaction rates for all steps of the catalytic reaction cycle, which by Arrhenius' law is given as the exponential $\exp(-\Delta E/k_B T)$ of the activation energy $\Delta E$. Highly accurate energies are required for reliable predictions, and classical algorithms cannot always reach the required level of accuracy, because the scaling of classical algorithms to reach the highest level of accuracy is exponential. By contrast, quantum algorithms for the same problem only scale polynomially, which enables efficient and accurate calculations of activation energies that are classically intractable.

A noteworthy example of how quantum computer-enhanced computational catalysis could benefit humanity would be the local and small-scale development of ammonia-based fertilizer for food production. The existing Haber-Bosch process for ammonia production requires high temperatures and pressures and large chemical plants. As a result, it is not particularly accessible to lower income communities. Quantum computers could help researchers discover new catalysts and reaction pathways for ammonia production on a local scale, making fertilizers more accessible to communities most in need and thereby lead to increased food production.

More work must be done to identify impactful problems for quantum computers. These problems must be (1) economically impactful and scientifically relevant, (2) out of scope for classical computers, and (3) realistically solvable on future quantum computers. Development of these use cases will require global expertise and collaboration. The recently established Open Quantum Institute [11], in partnership with quantum researchers globally, is taking an important first step by publishing an annual study on how to apply quantum computing to the U.N. Sustainable Development Goals [12], including some of the applications mentioned above.

## The need for quantum supercomputers

The quantum industry is currently focused on noisy physical qubits in noisy intermediate scale quantum (NISQ) machines. Solving the chemistry and material science problems that underpin the U.N. Sustainable Development Goals will require advancement to scalable, fault-tolerant quantum machines: quantum supercomputers with hundreds to thousands of logical qubits.

The requirements to solve these problems can be estimated using the Azure Quantum Resource Estimator [13,14]. Quantum advantage for simulations of quenches in the quantum dynamics of small, yet classically intractable problems [17,18] can be run on machines with a couple hundred logical qubits. The simulation of properties of scientifically interesting quantum magnets and effective models for materials, such as the Hubbard model, requires more than a thousand logical qubits, and the above-mentioned chemistry problems require a similar number of qubits with fast clock speeds [13,15,16].

Translating the logical qubit requirements into physical estimates using the Azure Quantum Resource Estimator [13,17,18] demonstrates that quantum dynamics applications will require systems with a couple hundred logical qubits, corresponding to more than 30,000 physical qubits. Quantum chemistry will require more than a million fast qubits and weeks of computation time.

While the development of such scaled quantum supercomputers will take years, we can start addressing the most impactful problems in chemistry and material science today using advances in classical high-performance computing (HPC) and artificial intelligence (AI) [19]. At the same time, we must remain resolutely focused on developing a quantum supercomputer, one that is capable of solving the real-world chemistry and material science problems that will directly address our most immediate global challenges.

# Use: Ensure that quantum machines are used responsibly and protect against malicious use

Inherent in the development of any transformative technology is the risk it will be used for harm. Quantum computing is no exception, and we cannot minimize or undersell the risk new cryptanalysis capabilities could have on cybersecurity. Nor can we ignore the ways dual-use risks will materialize in the application of classical and quantum computing to chemistry. These realities make it vitally important that industry, governments, and academia act collectively to steer this new technology toward societally beneficial outcomes.

## Understanding the risk to cryptography

Perhaps the best-known and well-studied risk associated with quantum computing is that a malicious actor could use future quantum supercomputers to break current asymmetric or public key cryptography, like RSA or ECC, by running Shor's algorithm [3]. Like the chemistry applications discussed above, Shor's algorithm run on a quantum supercomputer would offer a super-polynomial speedup, leading to practical quantum advantage over classical supercomputing.

A quantum machine capable of meaningfully threatening existing asymmetric encryption using Shor's algorithm would require more than one million qubits [13], which is many orders of magnitude beyond the power of current quantum machines. However, the present risk is a "store now, decrypt later" scenario, where a malicious actor stores encrypted data now (or has been doing so for years) with the expectation they will be able to decrypt the data once a quantum supercomputer is available. This risk presents an urgent need to quickly adopt quantum-safe security measures to limit the amount of future data that could be decrypted. Attention should be paid to systems and data with a security lifetime of many years, including systems related to national security, critical infrastructure, and personal or health-related data.

Whereas quantum computers could leverage Shor's algorithm to effectively break RSA or ECC, the only known threat vector against popular symmetric key algorithms, such as AES, is Grover's algorithm [20,21]. Grover's algorithm offers only a quadratic speedup, unlike Shor's super-polynomial speedup, and will therefore not present practical quantum advantage. More concretely, resource estimation research demonstrates Grover's algorithm on standard key sizes of AES has a runtime longer than the age of the universe for even the most optimistic assumptions of future qubit technology [22]. This research aligns with U.S. government views that AES symmetric key cryptography is resistant to quantum computers even without increasing the key size [23].

## Charting the path to a quantum-safe future

Mitigation of this quantum cybersecurity risk requires global migration to quantum-safe cryptographic systems. Where possible, transitioning away from public key to symmetric key encryption should begin today. Where this is not practical, the adoption of post-quantum cryptography (PQC) is necessary.

Public institutions and private industry have researched the threat to asymmetric, public key cryptography and the best mitigation approaches for decades. In the United States, for example, the National Institute of Standards and Technology (NIST) started testing and developing new quantum-resistant cryptographic standards in 2016 and plans to release the first set of new standards in 2024. Standardization efforts outside the U.S. are also underway in the EU [23], China [25], and with international standards bodies, including the ISO [26] and the IETF [27]. Given the magnitude of effort the PQC migration will require and the present risk of "store now, decrypt later," governments and industry must immediately prioritize and fund the transition. Efforts are already underway, but it is imperative that they be accelerated.

Quantum-key distribution (QKD) is a more nascent quantum-safe technology that cannot be considered secure on its own. QKD relies on the principles of quantum mechanics to securely exchange cryptographic keys with theoretically provable security guarantees. However, its current technical limitations do not achieve this guarantee and the lack of authentication protocols requires its use in hybrid form together with, and not instead of, PQC deployment [28].

Migration to quantum-safe security measures is the most meaningful proactive risk management strategy industry and governments can apply. As a complementary measure, quantum computing providers should adopt reactive safeguards that prevent cryptanalysis applications on future quantum hardware by implementing technical measures and processes that screen applications run on quantum computers. Similar to current virus protection, safeguards must include automated or manual review of the execution of user code. One approach is to offer only SaaS cloud access to quantum hardware for chemistry applications. These limited use application offerings would aim to give access only to users intending to apply quantum computing power to beneficial chemistry use cases. However, as discussed in the following section, the dual-use nature of chemistry research and development (R&D) means additional customer-based controls must be applied to ensure a user is not using this technology maliciously.

## Dual-use of scientific advancements & the need for responsible computing

For those driving advancements in chemistry, dual-use is a centuries-old challenge, illustrated throughout history and most clearly with Alfred Nobel's creation of dynamite. To Nobel's distress, dynamite—initially created for mining and construction in an era of great industrial growth—was quickly repurposed for warfare [29]. Today, technological developments in computational chemistry leave open the possibility that, rather than promoting societally beneficial applications, an actor with malicious intent could use these new technologies for harmful purposes [30,31]. Such concerns are not quantum-specific but are common to all technological advances in chemistry R&D. Advances in generative AI and HPC for chemistry raise new dual-use risks that must be addressed today.

Dual-use risk in chemistry applications with generative AI models was studied by researchers who demonstrated how generative AI could be used to produce *de novo* chemical threats [32]. This study examined the impact of inverting a machine learning model meant to avoid toxicity for consumer and pharmaceutical applications and instead reward predicted toxicity. As noted in a later paper, thinking around risks and mitigations associated with advanced computing and chemistry is still developing [33]. This is also true for regulation related to the use of AI models in chemistry, where initial efforts to define controls for AI models that could be exploited to develop chemical, biological, radiological, or nuclear (CBRN) weapons [34] demonstrate the need for partnership in building frameworks and controls for this technology [33].

Building on the above recommendations, industry and academic leaders, in collaboration with governments, should (1) drive awareness of science-based risks related to disruptive technologies and (2) implement meaningful governance and risk management measures to address these risks.

Industry must also leverage best practices from other dual-use computing applications. Specifically, the private sector should develop meaningful "know your customer" processes that aim to discern user intent. These measures are among the best tools currently available at reducing the likelihood of misuse by bad actors. This focus on controls aimed to manage "end-users and end-uses" [35,36] is an approach leading industry players [37,38], think tanks [39], and policymakers [34] have adopted and continue to advocate for around generative AI. Applying this approach first to the use of generative AI and HPC in chemistry and eventually to quantum computing would reduce the risk of a malicious actor gaining access to this disruptive computing power.

We must also convene across government, industry, academia, and civil society to craft a collective response to these risks. Frameworks for the development of "responsible quantum computing" or "responsible quantum technology" have begun appearing, particularly in academic literature [40]. International bodies and multinational forums, such as the OECD (Organization for Economic Co-operation and Development) [41] are also beginning to point to the importance of "responsible quantum technology." While these conversations must continue, organizations and researchers should shift to address "responsible computing" more broadly, rather than focusing specifically on "responsible quantum computing." This would allow the ecosystem to not only address computing risks holistically—representing the growing power of both classical and quantum computing—but also to be more theoretically congruent. "Responsible quantum computing or technology" implies that we are solving for risks that will not exist until quantum supercomputing exists in the future. For risks that are quantum-specific, the immediate mitigation path is clear: industry and governments must deploy quantum-safe encryption today. To mitigate dual-use risks in computational chemistry that have already emerged, stakeholders must partner to build harmonized "responsible computing" controls that scale as compute power advances through HPC, AI, and eventually to quantum supercomputing.

Learned experience with responsible AI has shown that early adoption and use of risk mitigation measures sets a crucial foundation for future innovation. Indeed, in 2018, Microsoft published the first version of the Responsible AI principles [42], years before the emergence of GPT-3 in Microsoft products. As the ecosystem thinks about how to develop "responsible computing" governance and accountability practices for the next generation of dual-use technology, 155 years since the invention of Nobel's dynamite [43], we should reflect on learned lessons from other scientific and ethical contexts. As an industry, we must ask ourselves: how can we create the right protections to avoid causing harm? How will we ensure we enable the positive, global use of computing while reducing the chance a bad actor could use this technology for malicious purposes? Together, as a global ecosystem, we in industry, government, and academia must ask ourselves: once we build this technology, how do we want it to be used?

# **Access:** Democratize the potential for economic growth across all of society

For quantum computing to genuinely deliver on its promise, we must enable equitable access to its benefits. Achieving this outcome will require a multi-pronged approach. In particular, it will require expanding the number of countries with capacity to develop and adopt quantum computing capabilities and the applications it enables, building workforce numbers and skills, and investing in foundational digital infrastructure. Industry, governments, academia, and civil society must take coordinated, directed action to achieve these outcomes, because now is the time we are best positioned to expand quantum computing's future benefits across all of society.

To ensure the equitable impact of quantum computing globally, governments, in close partnership with industry and academic partners, should focus on building the national capacity of countries that have been historically underrepresented in technological development. The majority of government quantum investment and quantum start-up incubation is currently concentrated in a small number of countries. There are only approximately thirty national quantum programs and thirty-eight countries represented in the quantum start-up ecosystem [44,45]. Of existing national funding for quantum programs, approximately 10% is in the United States, 33% is in Europe, and 39% is in China [44]. Industry must work carefully with governments and civil society to ensure that this technology is not built and adopted exclusively by select portions of the world or select sections of the economy. In 2023, the Geneva Science and Diplomacy Anticipator (GESDA) initiated the Open Quantum Institute with the objective of ensuring "future quantum computing is used for the common good" [46] to start making progress on this goal.

Industry and government must also empower a global workforce that can effectively work across borders to drive quantum computing's technological development as well as the deployment of beneficial quantum applications. Currently, the quantum computing industry is constrained by a shortage of talent, and the relatively small workforce that does exist is dispersed across nationally funded research institutions, private research institutions, multinational companies, and startups. In countries where quantum information science has become a national strategic priority, governments have sought to fund and expand national workforce and skilling initiatives to enable the development of quantum hardware and software. These programs must continue in order to accelerate technological development.

In addition, however, governments should also invest in new education and skilling efforts to equip a future workforce that will actually leverage quantum computing in its work—particularly in chemistry and material science. Initiatives to build a quantum-ready workforce should target fields related to advanced computing for chemistry and material science. Furthermore, they should establish and fund pathways from basic STEM education to terminal degrees and include upskilling opportunities for current workforce members. Diversifying the global workforce to prioritize skilling on quantum applications and impact, rather than only on the hardware and software itself, will lower barriers to entry for the adoption of quantum supercomputing when it arrives.

Finally, governments must invest in building the underlying digital infrastructure necessary for economic acceleration. Specifically, they should focus on ensuring access to the cloud, broadband, and 5G and other advanced mobile telecommunications. The AI era has taught us a clear lesson: countries must prepare so that they can embrace new technology as it emerges. Digital infrastructure investments are essential for such preparation. Not only will they enable continued progress of quantum technologies but also the creation of a more equitable quantum ecosystem.

Quantum computing offers transformative potential to advance solutions for some of society's most pressing issues. But the existence of qubits alone will not deliver these results. Rather, industry, governments, and academia must act in concert to tackle the impact, use, and access issues this new technology presents. We have reached a point where we understand how to mitigate quantum computing's attendant security risks; we must now channel it toward societally positive applications. Deep partnership and proactive collaboration will enable us to effectively harness the full power of this profound new tool.

Author contact information:

Matthias Troyer (matthias.troyer@microsoft.com)

Emily Violi Benjamin (emilyvioli@microsoft.com)

Ani Gevorkian (anigevorkian@microsoft.com)

32. F. Urbina, F. Lentzos, C. Invernizzi, and S. Ekins, "Dual use of artificial-intelligence-powered drug discovery," Nature Machine Intelligence, vol. 4, pp. 189-191, 2022. [Online].
Available: https://doi.org/10.1038/s42256-022-00465-9
33. F. Urbina, F. Lentzos, C. Invernizzi, and S. Ekins, "Preventing AI from creating biochemical threats," Journal of Chemical Information and Modeling, vol. 63, no. 3, pp. 691-694, 2023. [Online].
Available: https://doi.org/10.1021/acs.jcim.2c01616
34. The White House, "Executive Order on the Safe, Secure, and Trustworthy Development and Use of Artificial Intelligence," The White House, Oct. 30, 2023. [Online].
Available: https://www.whitehouse.gov/briefing-room/presidential-actions/2023/10/30/executive-order-on-the-safe-secure-and-trustworthy-development-and-use-of-artificial-intelligence/
35. Microsoft, "Microsoft and OpenAI partner to propose digital transformation of export controls," Microsoft On the Issues, Nov. 10, 2020. [Online]. Available: https://blogs.microsoft.com/on-the-issues/2020/11/10/openai-partnership-digital-export-controls/
36. U.S. Department of Commerce, "BIS-2018-0024-0175," Regulations.gov. [Online].
Available: https://www.regulations.gov/document/BIS-2018-0024-0175
37. Microsoft, "Governing AI: A Blueprint for the Future," [Online].
Available: https://query.prod.cms.rt.microsoft.com/cms/api/am/binary/RW14Gtw
38. OpenAI, "Safety Best Practices," OpenAI. [Online].
Available: https://platform.openai.com/docs/guides/safety-best-practices
39. RAND Corporation, "A model for regulating AI," RAND Corporation, Aug. 2023. [Online].
Available: https://www.rand.org/pubs/commentary/2023/08/a-model-for-regulating-ai.html
40. M. Kop, M. Aboy, E. De Jong, U. Gasser, T. Minssen, I. G. Cohen, M. Brongersma, T. Quintel, L. Floridi, and R. Laflamme, "Towards Responsible Quantum Technology," Harvard Berkman Klein Center for Internet & Society Research Publication Series #2023-1, Harvard University, 2023. [Online].
Available: https://cyber.harvard.edu/publication/2023/towards-responsible-quantum-technology
41. OECD, "Global Forum on Technology," OECD. [Online].
Available: https://www.oecd.org/digital/global-forum-on-technology/
42. Microsoft, "Responsible AI principles and approach," Microsoft AI. [Online].
Available: https://www.microsoft.com/en-us/ai/principles-and-approach
43. Science History Institute, "Alfred Nobel," [Online].
Available: https://www.sciencehistory.org/education/scientific-biographies/alfred-nobel
44. Qureca, "Overview of Quantum Initiatives Worldwide 2023," 2023. [Online].
Available: https://qureca.com/overview-of-quantum-initiatives-worldwide-2023/
45. McKinsey & Company, "Quantum technology sees record investments, progress on talent gap," [Online]. Available: https://shorturl.at/ostK9 [Accessed: 2023].
46. GESDA, "GESDA launches new Open Quantum Institute in Geneva with Swiss, CERN and UBS support," LinkedIn. [Online]. Available: https://www.linkedin.com/pulse/gesda-launches-new-open-quantum-institute-geneva-swiss-cern
15